\documentclass{natureprintstyle}

\usepackage{caption} 
\DeclareCaptionLabelSeparator{vbar}{ $\vert$ }
\usepackage{floatrow}
 \usepackage[label font={bf, small},labelformat = simple]{subfig}

\usepackage{stfloats}
\usepackage{float}
\usepackage{amsmath}
\usepackage{amssymb}
\usepackage{marvosym} 
\usepackage{multicol} 
\usepackage{siunitx}

\usepackage{tabularx} 
\newcolumntype{Y}{>{\centering\arraybackslash}X} 

\usepackage{diagbox} 
\usepackage{multirow} 
\usepackage{graphicx}

\usepackage{color}
\definecolor{extended}{cmyk}{0.50,0.90,0,0.35}
\usepackage{comment}

\usepackage{fancyhdr}
\usepackage{authoraftertitle}
\def\MyMaketitle{%
  \newpage\spacing{0.5}\setlength{\parskip}{3pt}%
    {\scalefont{3.0}\noindent\sloppy%
        \begin{flushleft}\bfseries\MyTitle\end{flushleft} \par}%
    {\scalefont{1.0}\noindent\sloppy \MyAuthor \par}%
}

\renewenvironment{affiliations}{%
    \let\olditem=\item
    \rule{\textwidth}{0.4pt}\\*[6pt]
    \raggedright
    \renewcommand\item[1][]{$^{\arabic{enumi}}$\stepcounter{enumi}}
    \setcounter{enumi}{1}%
    \setlength{\parindent}{0in}%
    \sffamily\sloppy%
    \scalefont{0.75}
    }{\let\item=\olditem}

\newfloat{affiliationsFloat}{b}{}[section] 

\newenvironment{myAuthors}{%
    \rule{\textwidth}{0.4pt}\\*[6pt]
    \let\olditem=\author
    \raggedright
    \renewcommand\author{}
    \setlength{\parindent}{0in}%
    \sffamily\sloppy%
    \scalefont{1.0}
    }{\let\author=\olditem}
    
\renewenvironment{abstract}{%
    \rule{\textwidth}{0.4pt}\\*[6pt]
    \setlength{\parindent}{0in}%
    \setlength{\parskip}{0in}%
    \scalefont{1.25}
        }{\par}

\makeatletter
\let\saved@includegraphics\includegraphics
\AtBeginDocument{\let\includegraphics\saved@includegraphics}
\makeatother

\title{Revealing the short-range structure of the "mirror nuclei" $^3$H and $^3$He}

\begin{document}

\widowpenalty=0
\clubpenalty=0
\onecolumn

\MyMaketitle

\begin{flushright}
\begin{minipage}{0.75\textwidth}
\raggedright
{\bf
\begin{myAuthors}
\author{S.~Li$^{1,2}$,}
\author{R.~Cruz-Torres$^{3,2}$,}
\author{N.~Santiesteban$^{1,3}$,}
\author{Z.~H.~Ye$^{4,5}$}
\author{D.~Abrams$^6$,}
\author{S.~Alsalmi$^{7,41}$,}
\author{D.~Androic$^8$,}
\author{K.~Aniol$^9$,}
\author{J.~Arrington$^{2,5,*}$,}
\author{T.~Averett$^{10}$,}
% ...

\end{myAuthors}
}

\vspace{12pt}

\begin{abstract}

When protons and neutrons (nucleons) are bound into atomic nuclei, they are close enough together to feel significant attraction, or repulsion, from the strong, short-distance part of the nucleon-nucleon interaction. These strong interactions lead to hard collisions between nucleons, generating pairs of highly-energetic nucleons referred to as short-range correlations (SRCs).
SRCs are an important but relatively poorly understood part of nuclear structure\cite{frankfurt88, sargsian03,Arrington:2022sov} and mapping out the strength and isospin structure (neutron-proton vs proton-proton pairs) of these virtual excitations is thus critical input for modeling a range of nuclear, particle, and astrophysics measurements~\cite{Hen:2016kwk, Arrington:2021alx, Arrington:2022sov}. Hitherto measurements used two-nucleon knockout or ``triple-coincidence'' reactions to measure the relative contribution of np- and pp-SRCs by knocking out a proton from the SRC and detecting its partner nucleon (proton or neutron). These measurements\cite{Subedi:2008zz, Korover:2014dma, CLAS:2018xvc} show that SRCs are almost exclusively np pairs, but had limited statistics and required large model-dependent final-state interaction (FSI) corrections. We report on the first measurement using inclusive scattering from the mirror nuclei $^3$H and $^3$He to extract the np/pp ratio of SRCs in the A=3 system. We obtain a measure of the np/pp SRC ratio that is an order of magnitude more precise than previous experiments, and find a dramatic deviation from the near-total np dominance observed in heavy nuclei. This result implies an unexpected structure in the high-momentum wavefunction for $^3$He and $^3$H. Understanding these results will improve our understanding of the short-range part of the N-N interaction. 

\end{abstract}

\end{minipage}
\end{flushright}
\begin{affiliationsFloat}
\begin{affiliations}
\item University of New Hampshire, Durham, New Hampshire 03824, USA
\item Lawrence Berkeley National Laboratory, Berkeley, California 94720, USA
\item Massachusetts Institute of Technology, Cambridge, Massachusetts 02139, USA
\item Tsinghua University, Beijing, China
\item Physics Division, Argonne National Laboratory, Lemont, Illinois 60439, USA
\item University of Virginia, Charlottesville, Virginia 22904, USA
\item Kent State University, Kent, Ohio 44240, USA
\item University of Zagreb, Zagreb, Croatia
\item California State University, Los Angeles, California 90032, USA
\item The College of William and Mary, Williamsburg, Virginia 23185, USA
\item University of Tennessee, Knoxville, Tennessee 37966, USA
\item INFN Sezione di Catania, Italy
\item Mississippi State University, Mississippi State, Mississippi 39762, USA
\item Hampton University, Hampton, Virginia 23669, USA
\item Old Dominion University, Norfolk, Virginia 23529, USA
\item Thomas Jefferson National Accelerator Facility, Newport News, Virginia 23606, USA
\item Florida International University, Miami, Florida 33199, USA
\item Michigan State University, East Lansing, Michigan 48824, USA
\item Stony Brook, State University of New York, New York 11794, USA
\item University of Connecticut, Storrs, Connecticut 06269, USA
\item INFN, Rome, Italy
\item Tohoku University, Sendai, Japan
\item Temple University, Philadelphia, Pennsylvania 19122, USA
\item Ohio University, Athens, Ohio 45701, USA
\item Nuclear Research Center -Negev, Beer-Sheva, Israel
\item Columbia University, New York, New York 10027, USA
\item University of Manitoba, Winnipeg, MB R3T 2N2, Canada
\item Jo\v{z}ef Stefan Institute, 1000 Ljubljana, Slovenia
\item Faculty of Mathematics and Physics, University of Ljubljana, 1000 Ljubljana, Slovenia
\item Institut f\"{u}r Kernphysik, Johannes Gutenberg-Universit\"{a}t Mainz, DE-55128 Mainz, Germany
\item Christopher Newport University, Newport News, Virginia 23606, USA
\item Saint Norbert College, De Pere, Wisconsin 54115, USA
\item Institute of Physics and Technology, Kharkov, Ukraine
\item Norfolk State University, Norfolk, Virginia 23529, USA
\item Rutgers University, New Brunswick, New Jersey 08854, USA
\item Divisi\'{o}n de Ciencias y Tecnolog\'{i}a, Universidad Ana G. M\'{e}ndez, Recinto de Cupey, San Juan 00926, Puerto Rico
\item Yerevan Physics Institute, Yerevan, Armenia
\item Syracuse University, Syracuse, New York 13244, USA
\item Northern Michigan University, Marquette, Michigan 49855, USA
\item University of Glasgow, Glasgow, G12 8QQ Scotland, UK
\item King Saud University, Riyadh 11451, Kingdom of Saudi Arabia
\hspace{1cm}
* email:JArrington@lbl.gov
\end{affiliations}
\end{affiliationsFloat}
\sloppy


\def\micro{\mu}
\def\deg{^\circ}
\def\gtorder{\mathrel{\raise.3ex\hbox{$>$}\mkern-14mu
 \lower0.6ex\hbox{$\sim$}}}
\def\ltorder{\mathrel{\raise.3ex\hbox{$<$}\mkern-14mu
 \lower0.6ex\hbox{$\sim$}}}
\def\etal{\textit{et al.}}

\clearpage
\Large      
\setlength{\lineskip}{3pt}
Nuclei are bound by the attractive components of the nucleon-nucleon (NN) interaction and the low-momentum part of their wavefunction is accurately described by mean-field or shell-model calculations\cite{kelly96}. These calculations show that the characteristic nucleon momenta in nuclei grow with target mass number A in light nuclei, becoming roughly constant in heavy nuclei. The strong, short-distance components of the NN interaction - the tensor attraction and short-range repulsive core - give rise to hard interactions between pairs of nucleons that are not well captured in mean-field calculations. These hard interactions create high-momentum nucleon pairs - two-nucleon short-range correlations (2N-SRCs) - which embody the universal two-body interaction at short distances and have a common structure in all nuclei\cite{frankfurt88, frankfurt93}.

SRCs are challenging to isolate in conventional low-energy measurements, but can be cleanly identified in inclusive electron scattering experiments for carefully chosen kinematics. Elastic electron-proton (e-p) scattering from a stationary nucleon corresponds to $x = Q^2 / (2M\nu) = 1$, where $Q^2$ is the four-momentum transfer squared, $\nu$ is the energy transfer, and $M$ is the mass of the proton. Scattering at fixed $Q^2$ but larger energy transfer ($x<1$) corresponds to inelastic scattering, where the proton is excited or broken apart. Scattering at lower energy transfer ($x>1$) is kinematically forbidden for a stationary proton, but larger $x$ values are accessible as the initial nucleon momentum increases, providing a way to isolate scattering from moving nucleons and thus study high-momentum nucleons in SRCs\cite{sargsian03, frankfurt93}.

Inclusive A(e,e$^\prime$) measurements at SLAC\cite{frankfurt93} and Jefferson Lab (JLab)\cite{fomin12, schmookler19} compared electron scattering from heavy nuclei to the deuteron for $x>1.4$ at $Q^2>1.4$~GeV$^2$, isolating scattering from nucleons above the Fermi momentum. They found identical cross sections up to a normalization factor, yielding a plateau in the A/$^2$H ratio for $x>1.4$, confirming the picture that high-momentum nucleons are generated within SRCs and exhibit identical two-body behavior in all nuclei. Using this technique, experiments have mapped out the contribution of SRCs for a range of light and heavy nuclei\cite{frankfurt93, egiyan03, fomin12, schmookler19}.

\begin{figure*}[htb]
        \includegraphics[width=89mm]{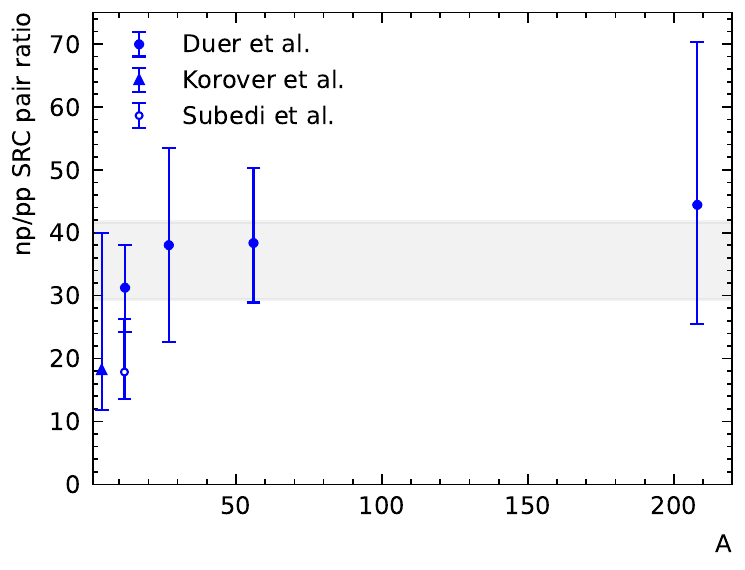}
        \caption{\textbf{Ratio of np-SRC to pp-SRCs in nuclei.} The ratio of np- to pp-SRC number from two-nucleon knockout measurements: solid circles\cite{CLAS:2018xvc}, solid triangle\cite{Korover:2014dma}, and hollow circle\cite{Subedi:2008zz}. Error bars indicate the 1$\sigma$ uncertainties, and the shaded band indicates the average ratio and 68\% confidence level region (excluding Ref.\cite{Subedi:2008zz} for which the FSI corrections applied are estimated to be $\sim$70\% too small\cite{CLAS:2018xvc}).}\label{fig:npoverpp-old}
\end{figure*}

Because inclusive A(e,e$^\prime$) scattering sums over proton and neutron knockout, it does not usually provide information on the isospin structure (np, pp, or nn) of these SRCs. The isospin structure has been studied using A(e,e$^\prime$pN$_s$) triple-coincidence measurements in which scattering from a high momentum proton is detected along with a spectator nucleon, N$_s$ (either proton or neutron), from the SRC pair with a momentum nearly equal but opposite to the initial proton. By detecting both np and pp final states, these measurements extract the ratio of np- to pp-SRCs and find that np-SRCs dominate\cite{Subedi:2008zz, Korover:2014dma, CLAS:2018xvc} while pp-SRCs have an almost negligible contribution, as seen in Fig.~\ref{fig:npoverpp-old}. 
Note that the observed np to pp ratio for SRCs depends somewhat on the range of nucleon momenta probed. This allows for measurements of the momentum dependence of the ratio\cite{Korover:2014dma}, but also means that direct comparisons of these ratios have to account for the momentum acceptance of each experiment. While these measurements provide unique sensitivity to the isospin structure, they have limited precision, typically 30--50\%, and require large final-state interaction (FSI) corrections.
Charge-exchange FSIs, where an outgoing neutron rescatters from one of the remaining protons in the nucleus, can produce a high-momentum proton in the final state that came from an initial state neutron (or vice versa). Because there are far more np-SRCs than pp-SRCs, even a small fraction of np pairs misidentified as pp will significantly modify the observed ratio\cite{Arrington:2022sov}. Modern calculations\cite{Colle:2015lyl} suggest that this nearly doubles the number of pp-SRCs detected in the final state\cite{CLAS:2018xvc}, while earlier analyses estimated a much smaller ($\sim$15\%) enhancement\cite{Subedi:2008zz}. Because of this, we exclude the data of Ref.\cite{Subedi:2008zz} in further discussion. Combining the remaining measurements in Fig.~\ref{fig:npoverpp-old}, we find that the average pp-SRCs is only (2.9$\pm$0.5)\% that of np-SRCs. This implies that the high-momentum tails of the nuclear momentum distribution is almost exclusively generated by np-SRCs and thus have nearly identical proton and neutron contributions, even for the most neutron rich nuclei.

This observed np dominance was shown to be a consequence of the short-distance tensor attraction\cite{schiavilla07, wiringa08, wiringa14}, which yields a significant enhancement of high-momentum isospin-0 np pairs. The isospin structure of 2N-SRCs determines the relative proton and neutron contributions at large momentum, impacting scattering measurements (including neutrino oscillation measurements), nuclear collisions, and sub-threshold particle production, making a clear understanding of the underlying physics critical in interpreting a range of key measurements\cite{Frankfurt:2008zv, Hen:2016kwk, Arrington:2021alx, Arrington:2022sov, Lu:2022ngd}. In addition, the observation of an unexpected correlation between the nuclear quark distribution functions\cite{seely09} and SRCs\cite{fomin12} in light nuclei suggested the possibility that they are driven by the same underlying physics. If so, the isospin structure of SRCs could translate into a quark flavor dependence in the nuclei. While this possibility has been examined in comparisons of EMC and SRC measurements\cite{arrington12b, schmookler19, arrington19, Arrington:2021vuu, Arrington:2022sov}, existing data are unable to determine if such a flavor dependence exists.

A new possibility for studying the isospin structure of SRCs was demonstrated recently when, for the first time, an inclusive measurement\cite{Nguyen:2020mgo} observed np-SRC enhancement by comparing the isospin distinct nuclei $^{48}$Ca and $^{40}$Ca. The measurement confirmed np-dominance, but only extracted a 68\% (95\%) confidence level upper limit on the pp/np ratio of 3.2\% (11.7\%). We report here the results of a significantly more precise extraction of the isospin structure of SRCs in the A = 3 system making use of the inclusive scattering from the mirror nuclei $^3$H and $^3$He. This avoids the large corrections associated with final-state interactions of the detected nucleons in two-nucleon knockout measurements, does not require a correction for the difference in mass between the two nuclei, and provides a dramatic increase in sensitivity compared to the measurements on calcium or previous two-nucleon knockout data.

Data for experiment E12-11-112 were taken in Hall A at JLab in 2018, covering the quasielastic (QE) scattering at $x\gtorder1$. Electrons were detected using two High Resolution Spectrometers (HRSs), described in detail in Ref.\cite{Alcorn:2004sb}, each consists of three focusing quadrupoles and one 45-degree dipole with a solid angle of $\sim$5~msr. The primary data were taken in the second run period (fall 2018) with a 4.332~GeV beam energy and the Left HRS at 17 degrees. This corresponds to $Q^2 \ge 1.4$~GeV$^2$ in the SRC plateau region, which has been demonstrated to be sufficient to isolate scattering from 2N-SRCs at large $x$\cite{frankfurt93, egiyan03, fomin17, Arrington:2022sov}. We also include data from experiment E12-14-011, taken during spring 2018 run period\cite{cruz-torres19} at 20.88$^\circ$ scattering angle, corresponding to $Q^2 \approx 1.9$~GeV$^2$ in the SRC plateau region. A new target system was developed for these experiments; details of the target system, including the first high-luminosity tritium target to be used in an electron scattering measurement in the last thirty years, are presented in the Methods section.

The electron trigger required signals from two scintillator planes and the CO$_2$ gas-Cherenkov chamber. Electron tracks were identified using the Cherenkov and two layers of lead-glass calorimeters, and reconstructed using two vertical drift chambers and optics matrices\cite{Alcorn:2004sb} were used to determine the angle, momentum, and position along the target for the scattered electrons. Acceptance cuts on the reconstructed scattering angle ($\pm$30~mrad in-plane, $\pm$60~mrad out-of-plane), momentum ($<$4\% from the central momentum), and target position (central 16~cm of the target). The final cut suppresses endcap contributions and the residual contamination was subtracted using measurements on an empty cell, as illustrated in Extended Data Fig.~\ref{fig:endcap}. The spectrometer acceptance was checked against Monte Carlo simulations and found to be essentially identical for all targets, so the cross section ratio is extracted from the yield ratio after after we apply a correction for the slight difference in the acceptance and radiative corrections. Additional details on the analysis and uncertainties is provided in the Methods section.

Meson-exchange currents (MEC) and isobar contributions are expected to be negligible\cite{sargsian03, arrington12a} for large energy transfers ($\nu \gtorder 0.5$~GeV), $Q^2 > 1$~GeV$^2$, and $x > 1$. To isolate SRCs, we take data with $x \ge 1.4$ and $Q^2>1.4$~GeV$^2$, which yields $\nu > 0.4$~GeV with an average value of 0.6~GeV. Final-state interactions at these kinematics are expected to be negligible\cite{sargsian03,arrington12a} except between the two nucleons in the SRC, and these are assumed cancel in the target ratios\cite{frankfurt88, sargsian03, Arrington:2022sov}. At $x>1$, the minimum initial momentum of the struck nucleon increases\cite{sargsian03} with $x$ and $Q^2$, and previous measurements have shown that for $Q^2 \ge 1.4$~GeV$^2$, $x>1.4$--1.5 is sufficient to virtually eliminate mean-field contributions and isolate 2N-SRCs. For the light nuclei considered here, scaling should be even more reliable:  the reduced Fermi momentum leads to a faster falloff of the mean-field contributions, providing earlier isolation of the SRCs, and any small residual MEC or FSI contributions (too small to see in previous A/$^2$H ratios) should have significant cancellation in the comparison of $^3$H to $^3$He. The radiative tail from the deuteron elastic contribution is subtracted and we excluded data as $x \to 2$ to avoid the rapid rise in the A/$^2$H ratios in the region where the deuteron cross section drops to zero.

\begin{figure*}[htb]
    \centering
            \subfloat[]{\includegraphics[width=89mm]{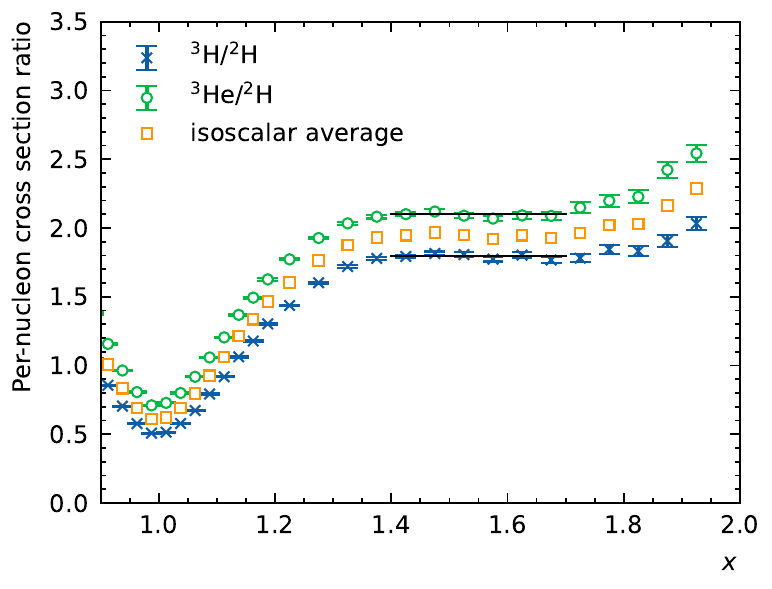}}
            \subfloat[]{\includegraphics[width=89mm]{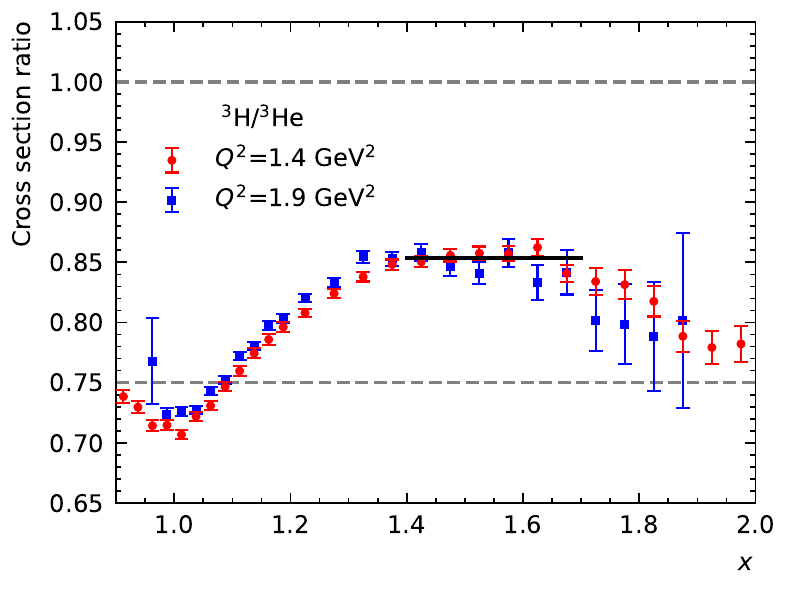}}
    \flushleft
        \caption{\textbf {Comparison of SRC contributions in $^3$He and $^3$H.} \textbf{a,} A/$^2$H per-nucleon cross section ratios for $^3$H, $^3$He, and ($^3$H+$^3$He)/2 from the $Q^2=1.4$~GeV$^2$ data. The solid lines indicate the combined $a_2$ value from the $Q^2=1.4$ and 1.9~GeV$^2$ data sets.  \textbf{b,} $R=^3$H/$^3$He cross section ratios vs. $x$; the $Q^2$ values are quoted for the SRC plateau region. The dashed lines indicate the predictions for np-dominance ($R=1$) and for isospin-independent SRCs ($R\approx0.75$).  For both figures, the error bars represent the combined statistical and uncorrelated systematic uncertainty (1$\sigma$ range); an additional normalization uncertainty of 1.18\% (0.78\% for the $^3$H/$^2$H ratio) is not shown.}\label{fig:nnratio_3_x2}
\end{figure*}

Fig.~\ref{fig:nnratio_3_x2}a shows the ratio of the cross section per nucleon from $^3$H and $^3$He to $^2$H from the $Q^2=1.4$~GeV$^2$ data set. The A/$^2$H ratio over the plateau region, $a_2(A)$, quantifies the relative contribution of SRCs in the nucleus A. We take $a_2$ to be the average for $1.4 \le x \le 1.7$ in this work, and combining the data from 1.4 and 1.9~GeV$^2$, we obtained $a_2 = 1.784\pm0.016$ for $^3$H and $a_2 = 2.088\pm0.026$ for $^3$He. The uncertainty includes the 0.78\% (1.18\%) uncertainty on the relative normalization of $^3$H ($^3$He) to $^2$H. We examined the impact of varying the $x$ region used to extract $a_2$ and for reasonable $x$ ranges, the cut dependence was negligible. Note that for $x>1.7$, there is an additional contribution from 2-body breakup in $^3$He relative to $^3$H, causing a deviation from the expected scaling in the SRC-dominated region\cite{CiofidegliAtti:2017xtx, Andreoli:2021cxo}. Because of this, we focus on $x<1.7$ where the comparison is not distorted by this contribution. A comparison of the $^3$He/$^2$H ratios at $Q^2$ of 1.4 and 1.9~GeV$^2$ to previous data is shown in Extended Data Fig.~\ref{fig:nnratio_D}, and all of the cross section ratios are given in Extended Data Tables~\ref{tab:data1} and~\ref{tab:data2}.

Fig.~\ref{fig:nnratio_3_x2}a also shows the unweighted average of the $^3$H/$^2$H and $^3$He/$^2$H ratios to provide $a_2$ for an ``isoscalar A = 3 nucleus''. We use the unweighted average of $a_2$ for $^3$H and $^3$He to avoid biasing the result towards the data set with smaller uncertainties. We also show a comparison of our two data sets to previous $^3$He/$^2$H ratios at higher $Q^2$ from JLab experiment E02-019\cite{fomin12} in Supplemental Fig.~\ref{fig:nnratio_D}. The results are in excellent agreement, with the onset of the plateau occurring slightly earlier in $x$ as $Q^2$ increases, as expected\cite{frankfurt93, fomin17, Arrington:2022sov}.

From isospin symmetry, we expect an identical number of np-SRCs for both nuclei with an additional pp(nn)-SRC contribution in $^3$He ($^3$H). Because the e-p elastic cross section is significantly larger than the e-n cross section, the $^3$He/$^2$H ratio in the SRC-dominated region will be larger than the $^3$H/$^2$H ratio if there is any contribution from pp-SRCs in $^3$He. A clearer way to highlight the contribution of pp-SRCs comes from a direct comparison of $^3$H and $^3$He, shown in Fig.~\ref{fig:nnratio_3_x2}b. While the ratios to the deuteron show a significant dip near $x=1$ due to the narrow QE peak for the deuteron, the fact that the momentum distribution is very similar for $^3$H and $^3$He yields a much smaller dip. The ratio in the SRC-dominated region is $0.854\pm0.010$ for $1.4<x<1.7$, including the normalization uncertainty, with negligible cut dependence.

If we take $^3$He ($^3$H) to contain $N_{np}$ np-SRC pairs and $N_{pp}$ pp-SRC (nn-SRC) pairs, based on the assumption of isospin symmetry for the mirror nuclei, and assume the cross section for scattering from the SRC is proportional to the sum of the elastic e-N scattering from the two nucleons, we obtain
\begin{equation}
\frac{\sigma_{^3H}}{\sigma_{^3He}} = \frac{1+\sigma_{p/n}+2R_{pp/np}}{1+\sigma_{p/n}(1+2R_{pp/np})} ~,  
\label{eq:final}
\end{equation}
where $\sigma_{p/n} = \sigma_{ep}/\sigma_{en}$ and the $R_{pp/np} = N_{pp}/N_{np}$. The full derivation, including a discussion of these assumptions, as well as small corrections applied to account for SRC motion in the nucleus, are included in the Methods section. Averaging over the 2N-SRC kinematics, we obtain $\sigma_{p/n}=2.47\pm0.05$ with the uncertainty including the range of $x$ and $Q^2$ of the measurement and the cross sections uncertainties. From Eq.~\ref{eq:final}, our measurement of $\sigma_{^3H}/\sigma_{^3He}$ gives $R_{pp/np}=0.228\pm0.022$. Accounting for the small difference between center-of-mass motion for different SRCs, as detailed in the Methods section,  we obtain $R_{pp/np} = 0.230\pm0.023$ - well below the simple pair-counting estimate of  $P_{pp/np}=0.5$ for $^3$He (only one pp pair, two possible np pairs), but also 10$\sigma$ above the assumption of total np-SRC dominance.

We also examine measurements of the $^3$He(e,e$^\prime$p)/$^3$H(e,e$^\prime$p) cross section ratio at large missing momenta ($P_m$) from the single nucleon knock-out experiment\cite{cruz-torres19} in a similar fashion. The average $^3$He/$^3$H cross section ratio for $250<P_m<400$~MeV/$c$ is $1.55\pm0.2$ after applying partial FSI corrections\cite{Sargsian:2004tz}. Taking the cross section at large $P_m$ to be proportional to the number of protons in SRCs, we obtain $R_{pp/np} = 0.28\pm0.10$ from the cross section ratios. The comparison of the $^3$He and $^3$H(e,e$^\prime$p) data to detailed calculations including FSI corrections except for charge-exchange contributions can be used to estimate the impact of charge exchange (see Fig.~3 of Ref.\cite{cruz-torres20}). This comparison suggests that FSI on the $^3$He/$^3$H ratio depends strongly on $P_m$, with a change of sign around 300--350~MeV/$c$, yielding significant cancellation in the $250 < P_m < 400$~MeV/$c$ range. Based on this estimate of the charge-exchange FSI\cite{cruz-torres20}, we assign an additional 10\% uncertainty associated with potential FSI effects, yielding a $^3$He/$^3$H ratio of $1.55\pm0.20\pm0.15$ and $R_{pp/np} = 0.28 \pm 0.13$, which we take as our extraction from the data of Ref.\cite{cruz-torres19}.

\begin{figure*}[htb]
    \centering
    \includegraphics[width=89mm]{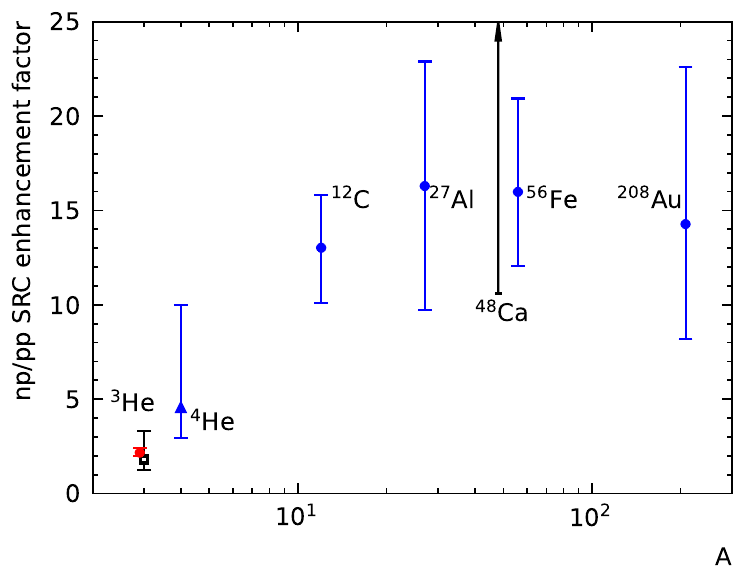}
    \flushleft
        \caption{\textbf {Nuclear enhancement of np-SRC over pp-SRCs. }
        Ratio of np-SRCs to pp-SRCs relative to the total number of np and pp pairs, for our new inclusive data (red circle), our extraction based on the proton knockout cross section ratios of Ref\cite{cruz-torres19} (black square), and previous two-nucleon knockout extractions\cite{Korover:2014dma, CLAS:2018xvc} (blue triangle, circles). For the $^{48}$Ca measurement\cite{Nguyen:2020mgo} (black arrow), we show the 68\% CL lower limit on the enhancement factor of 10.6; other points show the 1$\sigma$ uncertainty.}\label{fig:npoverpp}
\end{figure*}

To evaluate how much the np configuration is enhanced by the SRC mechanism, we compare the excess of the np-SRC/pp-SRC ratio ($R_{np/pp}$) over the pair-counting prediction $P_{np/pp} = (NZ)/(Z(Z-1)/2)$. Fig.~\ref{fig:npoverpp} shows this np-enhancement factor, $R_{np/pp}/P_{np/pp}$, from our new $^3$He/$^3$H inclusive data, our extraction from the $^3$He/$^3$H (e,e$^\prime$p) cross section ratios of Ref.\cite{cruz-torres19}, the published two-nucleon knockout measurements\cite{Korover:2014dma, CLAS:2018xvc}, and the inclusive measurement for $^{48}$Ca~\cite{Nguyen:2020mgo}. Note that for most nuclei shown in Fig.~\ref{fig:npoverpp-old}, $P_{np/pp} \approx 2$, while for $^4$He, $P_{np/pp}=4$, decreasing the $^4$He enhancement factor compared to those observed in heavier nuclei simply because of accounting for the available number of np and pp pairs.

Our new inclusive data yield $R_{np/pp}=4.34^{+.49}_{-.40}$, corresponding to an enhancement factor of $R_{np/pp}/P_{np/pp}=2.17^{+0.25}_{-0.20}$. Our extraction is significantly more precise than previous measurements and shows a clear deviation in $^3$He compared to heavy nuclei. Note that the different extractions of the np/pp ratios are not precisely equivalent, as there are small but important quantitative differences between the experiments and analyses. As discussed below, these differences do not appear to be responsible for the observed A dependence and may in fact be suppressing the true size of the difference.

While the np/pp extractions are often described as measuring the relative number of np- and pp-SRCs, they are more correctly described as the relative cross section contribution from SRCs over a specific range of initial nucleon momenta: $P_m$ of 250--400~MeV/$c$ for Ref.\cite{cruz-torres19}, 400--600~MeV/$c$ for Ref.\cite{Korover:2014dma} and 350--1000~MeV/$c$ for Ref.\cite{CLAS:2018xvc}. Both data\cite{Korover:2014dma} and calculations\cite{schiavilla07, wiringa08} suggest that the np/pp enhancement decreases at larger $P_m$ values, so if all exclusive measurements were examined in the same range, excluding the highest $P_m$ values, we would expect the enhancement to be even larger. Our inclusive measurement samples $P_m$ values of 250--300~MeV/$c$ and above, depending on the exact $x$, $Q^2$ bin, but yields a consistent cross section ratio for $1.4<x<1.7$ at both $Q^2$ values. While for lower $x$ and $Q^2$, the $P_m$ range extends below the coverage of the two-nucleon knockout measurements, the cross section at our larger $x$ values and $Q^2=1.9$~GeV$^2$ is dominated by $P_m \gtorder 350$~MeV/$c$, very similar to the exclusive measurements. In addition, for the $^3$He data, both our inclusive result and our extraction from the single-nucleon knockout\cite{cruz-torres19} data yield small enhancement factors, while the inclusive results on $^{48}$Ca, with very similar $P_m$ coverage, show a large enhancement, suggesting that the different missing momentum coverage is not responsible for the striking results in $^3$He.

One might speculate that the fact that the $^3$He has an extremely large deviation from N = Z might influence the isospin structure of the SRCs in some poorly understood way, but there are two reasons this seems unlikely to be the driving cause. First, the heaviest nuclei measured, $^{208}$Pb, also has a large proton-neutron asymmetry, N/Z = 1.54, but does not appear to have a significantly reduced enhancement factor. In addition, the $^4$He enhancement factor is also below all of the measurements on heavier nuclei, though the uncertainty does not allow us to make a definitive statement on its consistency with heavier nuclei. This points to the importance of making improved measurements of the np/pp SRC ratio, especially for light nuclei. While the measurement presented here yields dramatically smaller uncertainties, the technique requires nuclei with nearly identical structure but significant N/Z differences, so it cannot be applied widely. Even for other mirror nuclei, the sensitivity would be suppressed by a factor of $\Delta$Z/A, where $\Delta$Z is the difference in Z between the two nuclei.
Thus, improved measurements on $^4$He (or other light nuclei) will require two-nucleon knockout measurements with better statistics, possible at Jefferson Lab or the Electron-Ion Collider, as well as an improved understanding of the final-state interactions corrections.

The reduced np-SRC enhancement in $^3$He could also be related to the difference in the average nucleon separation in $^3$He compared to heavier nuclei. This would modify the relative importance of the different components of the NN potential. Therefore, this new measurement could be a way to constrain the relative contribution of the short-distance (isospin-dependent) tensor interaction and the very short-distance (isospin-independent) repulsive central core, which is difficult to constrain based on NN scattering data alone.

Finally, independent of the explanation for these surprising results, this measurement provides new insight into the high-momentum structure of $^3$He. The near-total np-SRC dominance seen in heavier nuclei suggested that the proton and neutron distributions would be essentially identical at large momenta, even for the extremely proton-rich $^3$He. Our new results suggests otherwise, indicating that the neutron plays a smaller role at high-momenta than if np-dominance is assumed, thus shifting strength between the high- and low-momentum regions. Because $^3$He plays an unique role as an effective polarized neutron target\cite{Blankleider:1983kb}, as well as recent extractions of the neutron structure function\cite{MARATHON:2021vqu}, a precise understanding of its microscopic structure is an key ingredient in a range of fundamental measurements in nuclear physics.

In conclusion, we have presented a novel measurement on the mirror nuclei $^3$H and $^3$He which provides a precise extraction of the enhancement of np-SRCs relative to pp-SRCs. The data show a significantly smaller enhancement of np-SRCs for A = 3 than seen in heavier nuclei, with uncertainties an order of magnitude smaller than previous two-nucleon knockout measurements. We also extracted the np/pp SRC ratio from $^3$He(e,e$^\prime$p)/$^3$H(e,e$^\prime$p) data\cite{cruz-torres19}, and found it to be consistent with the inclusive result, but with somewhat larger uncertainties. The new data on $^3$He, compared to heavier nuclei, suggests an unexpected and, as yet unexplained, A dependence in light nuclei. This surprising result makes available new information on the structure of these nuclei, which may impact a range of measurements that rely on understanding the $^3$H and $^3$He structure. These data may also play an important role constraining the relative contribution of the short-range attractive and repulsive parts of the nucleon-nucleon interaction.

\bibliography{tritium_SRC}

\begin{thebibliography}{10}
\expandafter\ifx\csname url\endcsname\relax
  \def\url#1{\texttt{#1}}\fi
\expandafter\ifx\csname urlprefix\endcsname\relax\def\urlprefix{URL }\fi
\providecommand{\bibinfo}[2]{#2}
\providecommand{\eprint}[2][]{\url{#2}}

\bibitem{frankfurt88}
\bibinfo{author}{Frankfurt, L.} \& \bibinfo{author}{Strikman, M.}
\newblock \bibinfo{title}{{Hard Nuclear Processes and Microscopic Nuclear
  Structure}}.
\newblock \emph{\bibinfo{journal}{Phys. Rept.}} \textbf{\bibinfo{volume}{160}},
  \bibinfo{pages}{235} (\bibinfo{year}{1988}).

\bibitem{sargsian03}
\bibinfo{author}{Sargsian, M.~M.} \emph{et~al.}
\newblock \bibinfo{title}{{Hadrons in the nuclear medium}}.
\newblock \emph{\bibinfo{journal}{J. Phys.}} \textbf{\bibinfo{volume}{G29}},
  \bibinfo{pages}{R1} (\bibinfo{year}{2003}).

\bibitem{Arrington:2022sov}
\bibinfo{author}{Arrington, J.}, \bibinfo{author}{Fomin, N.} \&
  \bibinfo{author}{Schmidt, A.}
\newblock \bibinfo{title}{{Progress in understanding short-range structure in
  nuclei: an experimental perspective}}.
\newblock \emph{\bibinfo{journal}{Ann. Rev. Nucl. Part. Sci.}}
  \textbf{\bibinfo{volume}{72}} (\bibinfo{year}{2022}).

\bibitem{Hen:2016kwk}
\bibinfo{author}{Hen, O.}, \bibinfo{author}{Miller, G.~A.},
  \bibinfo{author}{Piasetzky, E.} \& \bibinfo{author}{Weinstein, L.~B.}
\newblock \bibinfo{title}{{Nucleon-Nucleon Correlations, Short-lived
  Excitations, and the Quarks Within}}.
\newblock \emph{\bibinfo{journal}{Rev. Mod. Phys.}}
  \textbf{\bibinfo{volume}{89}}, \bibinfo{pages}{045002}
  (\bibinfo{year}{2017}).

\bibitem{Arrington:2021alx}
\bibinfo{author}{Arrington, J.} \emph{et~al.}
\newblock \bibinfo{title}{{Physics with CEBAF at 12 GeV and Future
  Opportunities}}.
\newblock \emph{\bibinfo{journal}{arXiv:2112.00060}}  (\bibinfo{year}{2021}).

\bibitem{Subedi:2008zz}
\bibinfo{author}{Subedi, R.} \emph{et~al.}
\newblock \bibinfo{title}{{Probing Cold Dense Nuclear Matter}}.
\newblock \emph{\bibinfo{journal}{Science}} \textbf{\bibinfo{volume}{320}},
  \bibinfo{pages}{1476} (\bibinfo{year}{2008}).

\bibitem{Korover:2014dma}
\bibinfo{author}{Korover, I.} \emph{et~al.}
\newblock \bibinfo{title}{{Probing the Repulsive Core of the Nucleon-Nucleon
  Interaction via the 4He(e,e$^\prime$pN) Triple-Coincidence Reaction}}.
\newblock \emph{\bibinfo{journal}{Phys. Rev. Lett.}}
  \textbf{\bibinfo{volume}{113}}, \bibinfo{pages}{022501}
  (\bibinfo{year}{2014}).

\bibitem{CLAS:2018xvc}
\bibinfo{author}{Duer, M.} \emph{et~al.}
\newblock \bibinfo{title}{{Direct Observation of Proton-Neutron Short-Range
  Correlation Dominance in Heavy Nuclei}}.
\newblock \emph{\bibinfo{journal}{Phys. Rev. Lett.}}
  \textbf{\bibinfo{volume}{122}}, \bibinfo{pages}{172502}
  (\bibinfo{year}{2019}).

\bibitem{kelly96}
\bibinfo{author}{Kelly, J.}
\newblock \bibinfo{title}{Nucleon knockout by intermediate-energy electrons}.
\newblock \emph{\bibinfo{journal}{Adv. Nucl. Phys.}}
  \textbf{\bibinfo{volume}{23}}, \bibinfo{pages}{75} (\bibinfo{year}{1996}).

\bibitem{frankfurt93}
\bibinfo{author}{Frankfurt, L.~L.}, \bibinfo{author}{Strikman, M.~I.},
  \bibinfo{author}{Day, D.~B.} \& \bibinfo{author}{Sargsyan, M.}
\newblock \bibinfo{title}{{Evidence for short-range correlations from high
  $Q^2$ (e,e') reactions}}.
\newblock \emph{\bibinfo{journal}{Phys. Rev. C}} \textbf{\bibinfo{volume}{48}},
  \bibinfo{pages}{2451} (\bibinfo{year}{1993}).

\bibitem{fomin12}
\bibinfo{author}{Fomin, N.} \emph{et~al.}
\newblock \bibinfo{title}{{New measurements of high-momentum nucleons and
  short-range structures in nuclei}}.
\newblock \emph{\bibinfo{journal}{Phys. Rev. Lett.}}
  \textbf{\bibinfo{volume}{108}}, \bibinfo{pages}{092502}
  (\bibinfo{year}{2012}).

\bibitem{schmookler19}
\bibinfo{author}{Schmookler, B.} \emph{et~al.}
\newblock \bibinfo{title}{{Modified structure of protons and neutrons in
  correlated pairs}}.
\newblock \emph{\bibinfo{journal}{Nature}} \textbf{\bibinfo{volume}{566}},
  \bibinfo{pages}{354} (\bibinfo{year}{2019}).

\bibitem{egiyan03}
\bibinfo{author}{Egiyan, K.~S.} \emph{et~al.}
\newblock \bibinfo{title}{{Observation of nuclear scaling in the A(e, e-prime)
  reaction at x(B) greater than 1}}.
\newblock \emph{\bibinfo{journal}{Phys. Rev.}} \textbf{\bibinfo{volume}{C 68}},
  \bibinfo{pages}{014313} (\bibinfo{year}{2003}).

\bibitem{Colle:2015lyl}
\bibinfo{author}{Colle, C.}, \bibinfo{author}{Cosyn, W.} \&
  \bibinfo{author}{Ryckebusch, J.}
\newblock \bibinfo{title}{{Final-state interactions in two-nucleon knockout
  reactions}}.
\newblock \emph{\bibinfo{journal}{Phys. Rev. C}} \textbf{\bibinfo{volume}{93}},
  \bibinfo{pages}{034608} (\bibinfo{year}{2016}).

\bibitem{schiavilla07}
\bibinfo{author}{Schiavilla, R.}, \bibinfo{author}{Wiringa, R.~B.},
  \bibinfo{author}{Pieper, S.~C.} \& \bibinfo{author}{Carlson, J.}
\newblock \bibinfo{title}{{Tensor Forces and the Ground-State Structure of
  Nuclei}}.
\newblock \emph{\bibinfo{journal}{Phys. Rev. Lett.}}
  \textbf{\bibinfo{volume}{98}}, \bibinfo{pages}{132501}
  (\bibinfo{year}{2007}).

\bibitem{wiringa08}
\bibinfo{author}{Wiringa, R.}, \bibinfo{author}{Schiavilla, R.},
  \bibinfo{author}{Pieper, S.} \& \bibinfo{author}{Carlson, J.}
\newblock \bibinfo{title}{{Dependence of two-nucleon momentum densities on
  total pair momentum}}.
\newblock \emph{\bibinfo{journal}{Phys. Rev.}} \textbf{\bibinfo{volume}{C 78}},
  \bibinfo{pages}{021001} (\bibinfo{year}{2008}).

\bibitem{wiringa14}
\bibinfo{author}{Wiringa, R.}, \bibinfo{author}{Schiavilla, R.},
  \bibinfo{author}{Pieper, S.} \& \bibinfo{author}{Carlson, J.}
\newblock \bibinfo{title}{{Nucleon and nucleon-pair momentum distributions in
  $A \le 12$ nuclei}}.
\newblock \emph{\bibinfo{journal}{Phys. Rev.}} \textbf{\bibinfo{volume}{C 89}},
  \bibinfo{pages}{024305} (\bibinfo{year}{2014}).

\bibitem{Frankfurt:2008zv}
\bibinfo{author}{Frankfurt, L.}, \bibinfo{author}{Sargsian, M.} \&
  \bibinfo{author}{Strikman, M.}
\newblock \bibinfo{title}{{Recent observation of short range nucleon
  correlations in nuclei and their implications for the structure of nuclei and
  neutron stars}}.
\newblock \emph{\bibinfo{journal}{Int. J. Mod. Phys. A}}
  \textbf{\bibinfo{volume}{23}}, \bibinfo{pages}{2991} (\bibinfo{year}{2008}).

\bibitem{Lu:2022ngd}
\bibinfo{author}{Lu, H.}, \bibinfo{author}{Ren, Z.} \& \bibinfo{author}{Bai,
  D.}
\newblock \bibinfo{title}{{Neutron-neutron short-range correlations and their
  impacts on neutron stars}}.
\newblock \emph{\bibinfo{journal}{Nucl. Phys. A}}
  \textbf{\bibinfo{volume}{1021}}, \bibinfo{pages}{122408}
  (\bibinfo{year}{2022}).

\bibitem{seely09}
\bibinfo{author}{Seely, J.} \emph{et~al.}
\newblock \bibinfo{title}{{New measurements of the EMC effect in very light
  nuclei}}.
\newblock \emph{\bibinfo{journal}{Phys. Rev. Lett.}}
  \textbf{\bibinfo{volume}{103}}, \bibinfo{pages}{202301}
  (\bibinfo{year}{2009}).

\bibitem{arrington12b}
\bibinfo{author}{Arrington, J.} \emph{et~al.}
\newblock \bibinfo{title}{{A detailed study of the nuclear dependence of the
  EMC effect and short-range correlations}}.
\newblock \emph{\bibinfo{journal}{Phys. Rev.}} \textbf{\bibinfo{volume}{C 86}},
  \bibinfo{pages}{065204} (\bibinfo{year}{2012}).

\bibitem{arrington19}
\bibinfo{author}{Arrington, J.} \& \bibinfo{author}{Fomin, N.}
\newblock \bibinfo{title}{{Searching for flavor dependence in nuclear quark
  behavior}}.
\newblock \emph{\bibinfo{journal}{Phys. Rev. Lett.}}
  \textbf{\bibinfo{volume}{123}}, \bibinfo{pages}{042501}
  (\bibinfo{year}{2019}).

\bibitem{Arrington:2021vuu}
\bibinfo{author}{Arrington, J.} \emph{et~al.}
\newblock \bibinfo{title}{{Measurement of the EMC effect in light and heavy
  nuclei}}.
\newblock \emph{\bibinfo{journal}{Phys. Rev. C}}
  \textbf{\bibinfo{volume}{104}}, \bibinfo{pages}{065203}
  (\bibinfo{year}{2021}).

\bibitem{Nguyen:2020mgo}
\bibinfo{author}{Nguyen, D.} \emph{et~al.}
\newblock \bibinfo{title}{{Novel observation of isospin structure of
  short-range correlations in calcium isotopes}}.
\newblock \emph{\bibinfo{journal}{Phys. Rev. C}}
  \textbf{\bibinfo{volume}{102}}, \bibinfo{pages}{064004}
  (\bibinfo{year}{2020}).

\bibitem{Alcorn:2004sb}
\bibinfo{author}{Alcorn, J.} \emph{et~al.}
\newblock \bibinfo{title}{{Basic Instrumentation for Hall A at Jefferson Lab}}.
\newblock \emph{\bibinfo{journal}{Nucl. Instrum. Meth. A}}
  \textbf{\bibinfo{volume}{522}}, \bibinfo{pages}{294} (\bibinfo{year}{2004}).

\bibitem{fomin17}
\bibinfo{author}{Fomin, N.}, \bibinfo{author}{Higinbotham, D.},
  \bibinfo{author}{Sargsian, M.} \& \bibinfo{author}{Solvignon, P.}
\newblock \bibinfo{title}{{New Results on Short-Range Correlations in Nuclei}}.
\newblock \emph{\bibinfo{journal}{Ann. Rev. Nucl. Part. Sci.}}
  \textbf{\bibinfo{volume}{67}}, \bibinfo{pages}{129} (\bibinfo{year}{2017}).

\bibitem{cruz-torres19}
\bibinfo{author}{Cruz-Torres, R.} \emph{et~al.}
\newblock \bibinfo{title}{{Comparing proton momentum distributions in $A=2$ and
  3 nuclei via $^2$H $^3$H and $^3$He $(e, e'p)$ measurements}}.
\newblock \emph{\bibinfo{journal}{Phys. Lett. B}}
  \textbf{\bibinfo{volume}{797}}, \bibinfo{pages}{134890}
  (\bibinfo{year}{2019}).

\bibitem{arrington12a}
\bibinfo{author}{Arrington, J.}, \bibinfo{author}{Higinbotham, D.},
  \bibinfo{author}{Rosner, G.} \& \bibinfo{author}{Sargsian, M.}
\newblock \bibinfo{title}{Hard probes of short-range nucleon–nucleon
  correlations}.
\newblock \emph{\bibinfo{journal}{Prog. Part. Nucl. Phys.}}
  \textbf{\bibinfo{volume}{67}}, \bibinfo{pages}{898} (\bibinfo{year}{2012}).

\bibitem{CiofidegliAtti:2017xtx}
\bibinfo{author}{Ciofi~degli Atti, C.} \& \bibinfo{author}{Morita, H.}
\newblock \bibinfo{title}{{Universality of many-body two-nucleon momentum
  distributions: Correlated nucleon spectral function of complex nuclei}}.
\newblock \emph{\bibinfo{journal}{Phys. Rev. C}} \textbf{\bibinfo{volume}{96}},
  \bibinfo{pages}{064317} (\bibinfo{year}{2017}).

\bibitem{Andreoli:2021cxo}
\bibinfo{author}{Andreoli, L.} \emph{et~al.}
\newblock \bibinfo{title}{{Electron scattering on A=3 nuclei from quantum Monte
  Carlo based approaches}}.
\newblock \emph{\bibinfo{journal}{Phys. Rev. C}}
  \textbf{\bibinfo{volume}{105}}, \bibinfo{pages}{014002}
  (\bibinfo{year}{2022}).

\bibitem{Sargsian:2004tz}
\bibinfo{author}{Sargsian, M.~M.}, \bibinfo{author}{Abrahamyan, T.~V.},
  \bibinfo{author}{Strikman, M.~I.} \& \bibinfo{author}{Frankfurt, L.~L.}
\newblock \bibinfo{title}{{Exclusive electrodisintegration of He-3 at high
  Q**2. I. Generalized eikonal approximation}}.
\newblock \emph{\bibinfo{journal}{Phys. Rev. C}} \textbf{\bibinfo{volume}{71}},
  \bibinfo{pages}{044614} (\bibinfo{year}{2005}).

\bibitem{cruz-torres20}
\bibinfo{author}{Cruz-Torres, R.} \emph{et~al.}
\newblock \bibinfo{title}{{Probing Few-Body Nuclear Dynamics via $^3$H and
  $^3$He(e,e'p)pn Cross-Section Measurements}}.
\newblock \emph{\bibinfo{journal}{Phys. Rev. Lett.}}
  \textbf{\bibinfo{volume}{124}}, \bibinfo{pages}{212501}
  (\bibinfo{year}{2020}).

\bibitem{Blankleider:1983kb}
\bibinfo{author}{Blankleider, B.} \& \bibinfo{author}{Woloshyn, R.~M.}
\newblock \bibinfo{title}{{Quasielastic Scattering of Polarized Electrons on
  Polarized $^{3}$He}}.
\newblock \emph{\bibinfo{journal}{Phys. Rev. C}} \textbf{\bibinfo{volume}{29}},
  \bibinfo{pages}{538} (\bibinfo{year}{1984}).

\bibitem{MARATHON:2021vqu}
\bibinfo{author}{Abrams, D.} \emph{et~al.}
\newblock \bibinfo{title}{{Measurement of the Nucleon $F^n_2/F^p_2$ Structure
  Function Ratio by the Jefferson Lab MARATHON Tritium/Helium-3 Deep Inelastic
  Scattering Experiment}}.
\newblock \emph{\bibinfo{journal}{Phys. Rev. Lett.}}
  \textbf{\bibinfo{volume}{128}}, \bibinfo{pages}{132003}
  (\bibinfo{year}{2022}).

\bibitem{target_NIM}
\bibinfo{author}{Brajuskovic, B.} \emph{et~al.}
\newblock \bibinfo{title}{{Thermomechanical design of a static gas target for
  electron accelerators}}.
\newblock \emph{\bibinfo{journal}{Nucl. Instrum. Meth. A}}
  \textbf{\bibinfo{volume}{729}}, \bibinfo{pages}{469} (\bibinfo{year}{2013}).

\bibitem{target_report}
\bibinfo{author}{Meekins, D.}
\newblock \bibinfo{title}{{Determination of solid and fluid target thickness
  from measurements, JLab Document Number: TGT-CALC-17-020}}
  (\bibinfo{year}{2020}).

\bibitem{Santiesteban:2018qwi}
\bibinfo{author}{Santiesteban, S.~N.} \emph{et~al.}
\newblock \bibinfo{title}{{Density Changes in Low Pressure Gas Targets for
  Electron Scattering Experiments}}.
\newblock \emph{\bibinfo{journal}{Nucl. Instrum. Meth. A}}
  \textbf{\bibinfo{volume}{940}}, \bibinfo{pages}{351} (\bibinfo{year}{2019}).

\bibitem{dasu_thesis}
\bibinfo{author}{Dasu, S.}
\newblock \emph{\bibinfo{title}{{Precision measurement of x, Q2 and
  A-Dependence of $R_L/R_T$ and F2 in deep inelastic scattering}}}.
\newblock Ph.D. thesis, \bibinfo{school}{University of Rochester}
  (\bibinfo{year}{1988}).

\bibitem{DeForest1983}
\bibinfo{author}{DeForest, T.}
\newblock \bibinfo{title}{Off-shell electron nucleon cross section. the impulse
  approximation}.
\newblock \emph{\bibinfo{journal}{Nucl. Phys.}}
  \textbf{\bibinfo{volume}{A392}}, \bibinfo{pages}{232} (\bibinfo{year}{1983}).

\bibitem{Arrington:2003qk}
\bibinfo{author}{Arrington, J.}
\newblock \bibinfo{title}{{Implications of the discrepancy between proton
  form-factor measurements}}.
\newblock \emph{\bibinfo{journal}{Phys. Rev. C}} \textbf{\bibinfo{volume}{69}},
  \bibinfo{pages}{022201} (\bibinfo{year}{2004}).

\bibitem{ye17}
\bibinfo{author}{Ye, Z.}, \bibinfo{author}{Arrington, J.},
  \bibinfo{author}{Hill, R.~J.} \& \bibinfo{author}{Lee, G.}
\newblock \bibinfo{title}{{Proton and Neutron Electromagnetic Form Factors and
  Uncertainties}}.
\newblock \emph{\bibinfo{journal}{Phys. Lett. B}}
  \textbf{\bibinfo{volume}{777}}, \bibinfo{pages}{8} (\bibinfo{year}{2018}).

\end{thebibliography}

\newpage

\textbf{METHODS:}

\textbf{Target details.}
A special target system was built to meet the goals of the tritium rungroup experiments\cite{MARATHON:2021vqu, cruz-torres19} while satisfying all safety requirements for tritium handling\cite{target_NIM}. Four identical aluminum cells, 25.00~cm long and 1.27~cm in diameter, contained gaseous deuterium, hydrogen, helium-3 and tritium, with areal densities of 142.2, 70.8, 53.2, and 84.8~mg/cm$^2$ (85.0~mg/cm$^2$ for the spring data taking on tritium) at room temperature\cite{target_report}. A fifth empty cell was used for background measurements. Before each run period, JLab sent an empty cell to Savannah River Site for the tritium filling; all other targets were prepared locally. 

The tritium in the target cell decays into $^3$He with a half-life of 12.3 years, yielding an average 4.0\% (1.2\%) $^3$H density reduction, and corresponding $^3$He contamination, for the first (second) run period. The $^3$H data were corrected using $^3$He data taken at the same settings. During the second run period ($Q^2 = 1.4$~GeV$^2$ data), we observed a narrow peak at $x=1$ in all tritium data. With low $Q^2$ calibration runs, we confirmed that the shape was consistent with scattering from hydrogen. Since the tritium fill data reports no hydrogen component\cite{target_report}, the best hypothesis for this hydrogen contamination is the residual water from the target filling followed by the H2O + T2 $\to$ 2HTO + H2 reaction. The observed hydrogen contamination requires 4.1\% of tritium gas in the tritium cell to have exchanged with hydrogen in the water to form HTO, which freezes on the target wall and so is removed from the effective target thickness. Note that beam heating effects would drive away any HTO that freezes on the target endcaps, and so the frozen HTO will not interact with the beam, and the hydrogen gas only contributes at $x \le 1$, so neither of these are a source of background events in the range of interest for the SRC studies presented here. However, the clear hydrogen elastic peak at $x=1$ allows us to determine the amount of hydrogen gas in the target, and hence the tritium lost to HTO, yielding a correction to the tritium target thickness of $4.1\pm0.2$\%.

\textbf{Data taking and analysis.}
During data-taking, the electron beam was limited to 22.5~$\mu$A and rastered to a 2$\times$2~mm$^2$ square to avoid damage to the target. Detailed descriptions of the raster and additional beamline instrumentation can be found in Ref.\cite{Alcorn:2004sb}.  The target gas is heated by the beam, quickly reaching an equilibrium state with a reduced gas density along the beam path. A detailed study of both the single-target yield and target-ratio as a function of beam current\cite{Santiesteban:2018qwi} shows that the tritium, deuterium, and helium-3 densities as seen by the beam decreased by 9.72\%, 9.04\%, and 6.18\%, respectively, at 22.5~$\mu$A. This effect is linear at low current with deviations from linearity at higher currents. A direct analysis of the yield ratios between different targets was also performed, yielding smaller corrections that are more linear with current. Based on this analysis, we apply a 0.2\% normalization uncertainty to the target ratios.
 
The trigger and detector efficiencies ($>99$\% for all runs) were measured and applied on a run-by-run basis, with the trigger efficiency determined using samples of events with looser triggers (requiring only one scintillator plane or no Cherenkov signal). Comparisons of the acceptance for the gas targets showed no visible difference, and uncertainties were estimated by examining the cut dependence of the acceptance-corrected yield ratios. Based on this we assign a 0.2\% normalization uncertainty and a 0.2\% uncorrelated uncertainty up to $x=1.7$; above this the statistical precision of this test was limited and we apply a 1\% uncorrelated uncertainty. Subtraction of the residual endcap contribution yields a 1--4\% correction, with an uncorrelated uncertainty equal to one-tenth of the correction applied to each $x$ bin and a normalization uncertainty taken to be 0.2\%.

The radiative corrections were calculated for both targets following the prescription of Ref.~\cite{dasu_thesis} and the yield ratios are corrected for the difference in these effects. We take a 0.3\% normalization and 0.2\% uncorrelated uncertainty associated with the uncertainty in the radiative correction procedure. The room-temperature target thickness uncertainty associated with the uncertainty of the temperature and pressure measurements along with the equation of state was 1\% for $^3$He and 0.4\% for the hydrogen isotopes. This is combined with the 0.2\% normalization uncertainty associated with beam heating effects (described above).  
Combining these uncertainties, we find uncorrelated uncertainties of 0.3-0.6\% in the target ratios in the SRC-dominated kinematics and a normalization uncertainty of 0.78\% for $^3$H/$^2$H ratios and 1.18\% for $^3$He/$^3$H or $^3$He/$^2$H.

\textbf{Details of the np/pp extraction.}
We begin by assuming isospin symmetry for $^3$H and $^3$He, i.e. the proton distributions in $^3$H are identical to the neutron distributions in $^3$He and vice-versa. Under this assumption, if $^3$He ($^3$H) contains $N_{np}$ np-SRC pairs and $N_{pp}$ pp-SRC (nn-SRC) pairs, the cross section ratio will be
\begin{equation}
\frac{\sigma_{^3H}}{\sigma_{^3He}} = \frac{N_{np}\sigma_{np}+N_{pp}\sigma_{nn}}{N_{np}\sigma_{np}+N_{pp}\sigma_{pp}} ~,
\label{eq:ratio}
\end{equation}
where $\sigma_{NN}$ is the cross section for scattering from an NN-SRC. Assuming that the effect of SRC center-of-mass motion is identical for all SRCs in $^3$H and $^3$He, the inclusive cross section from 2N-SRCs in the SRC-dominated regime is proportional to the sum of quasielastic scattering from the nucleons in the correlated pair, i.e. $\sigma_{np}=\sigma_{ep}+\sigma_{en}$, $\sigma_{pp}=2\sigma_{ep}$, and $\sigma_{nn}=2\sigma_{en}$.
Eq.~\ref{eq:ratio} can be rewritten such that the target ratio depends only on the ratio of the off-shell elastic e-p to e-n cross section ratio, $\sigma_{p/n} = \sigma_{ep}/\sigma_{en}$ and the ratio $R_{pp/np} = N_{pp}/N_{np}$, yielding 
\begin{equation}
\frac{\sigma_{^3H}}{\sigma_{^3He}} = \frac{1+\sigma_{p/n}+2R_{pp/np}}{1+\sigma_{p/n}(1+2R_{pp/np})} ~.  
\label{eq:ratio2}
\end{equation}
as given in the main text. For a bound nucleon, $\sigma_{eN}$ is a function of both $x$ and $Q^2$. We use the deForest CC1 off-shell prescription\cite{DeForest1983}, the proton cross section fit from Ref.\cite{Arrington:2003qk} (without two-photon exchange corrections) and neutron form factors from Ref.\cite{ye17} to calculate $\sigma_{p/n}$.

Eq.~\ref{eq:ratio2} assumes isospin symmetry and an identical center-of-mass momentum distribution for np- and pp-SRC. We estimate corrections associated with violation of these assumptions using \textit{ab inito} Greens Function Monte Carlo calculations\cite{wiringa14} of the momentum distributions for protons and neutrons in $^3$H and $^3$He, which accounts for the isospin-symmetry violation arising from the Coulomb interaction. These calculations are used to estimate the difference between the np-SRC and pp-SRC momentum distributions in $^3$He, and the difference between the np-SRC momentum distributions between $^3$H and $^3$He. For the A = 3 system, we take the SRC momentum to be balanced by the spectator nucleon, for kinematics where this nucleon is not to be part of an SRC (i.e. taking $k \le k_{Fermi}$). We find typical SRC momenta of 120 MeV/$c$, with the momentum of np-SRCs in $^3$H is roughly 2~MeV/$c$ larger than for $^3$He, and pp(nn)-SRCs momenta are approximately 12 MeV/$c$ larger than np-SRCs within $^3$He ($^3$He). Using the smearing formalism of Ref.\cite{fomin12}, and assuming a 100\% uncertainty on the estimated corrections, we find that the increased smearing in $^3$H increases the $^3$H/$^3$He ratio by (0.4$\pm$0.4)\%, raising the extracted pp/np value by (2.5$\pm$2.5)\%, while the increased pp(nn) smearing directly lowers the extracted pp/np ratio by (2$\pm$2)\%. We apply these corrections to the extracted pp/np ratio to obtain the final corrected value for $R_{pp/np}$

\noindent{\bf Competing Interests:} The authors declare no competing interests.

\noindent{\bf Data Availability:} The raw data from this experiment were generated at the Thomas Jefferson National Accelerator Facility and are archived in the Jefferson Lab mass storage silo. Access to these data and relevant analysis codes can be facilitated by contacting the corresponding author.

\noindent{\bf Corresponding Author:} Correspondence should be addressed to J.~Arrington (JArrington@lbl.gov).

\noindent{\bf Acknowledgements:} 
We acknowledge useful discussions with Omar Benhar, Claudio Ciofi degli Atti, Wim Cosyn, Alessandro Lovato, Noemi Rocco, Misak Sargsian, Mark Strikman, and Robert Wiringa, and the contribution of the Jefferson Lab target group and technical staff for design and construction of the Tritium target and their support running this experiment.

This work was supported in part by the Department of Energy's Office of Science, Office of Nuclear Physics, under contracts DE-AC02-05CH11231, DE-FG02-88ER40410, DE-SC0014168, DE-FG02-96ER40950, the National Science Foundation, including grant NSF PHY-1714809, and DOE contract DE-AC05-06OR23177 under which JSA, LLC operates JLab.

\noindent\textbf{Author Contributions:} J.A, D.D., D.W.H., P.S., and Z.H.Y. were the experiment co-spokespersons. S.L., N.S., R.C-T., Z.H.Y., R.E.M., and F.H. made significant contributions the setup of the experiment and/or data analysis. R.J.H., D.M., and P.S. contributed to the design and operation of the tritium target. The full collaboration participated in the data collection and/or detector calibration and data analysis.

\setcounter{figure}{0}
\begin{figure*}[htb]
 \renewcommand\figurename{Extended Data Figure}
    \centering
        \includegraphics[width=89mm]{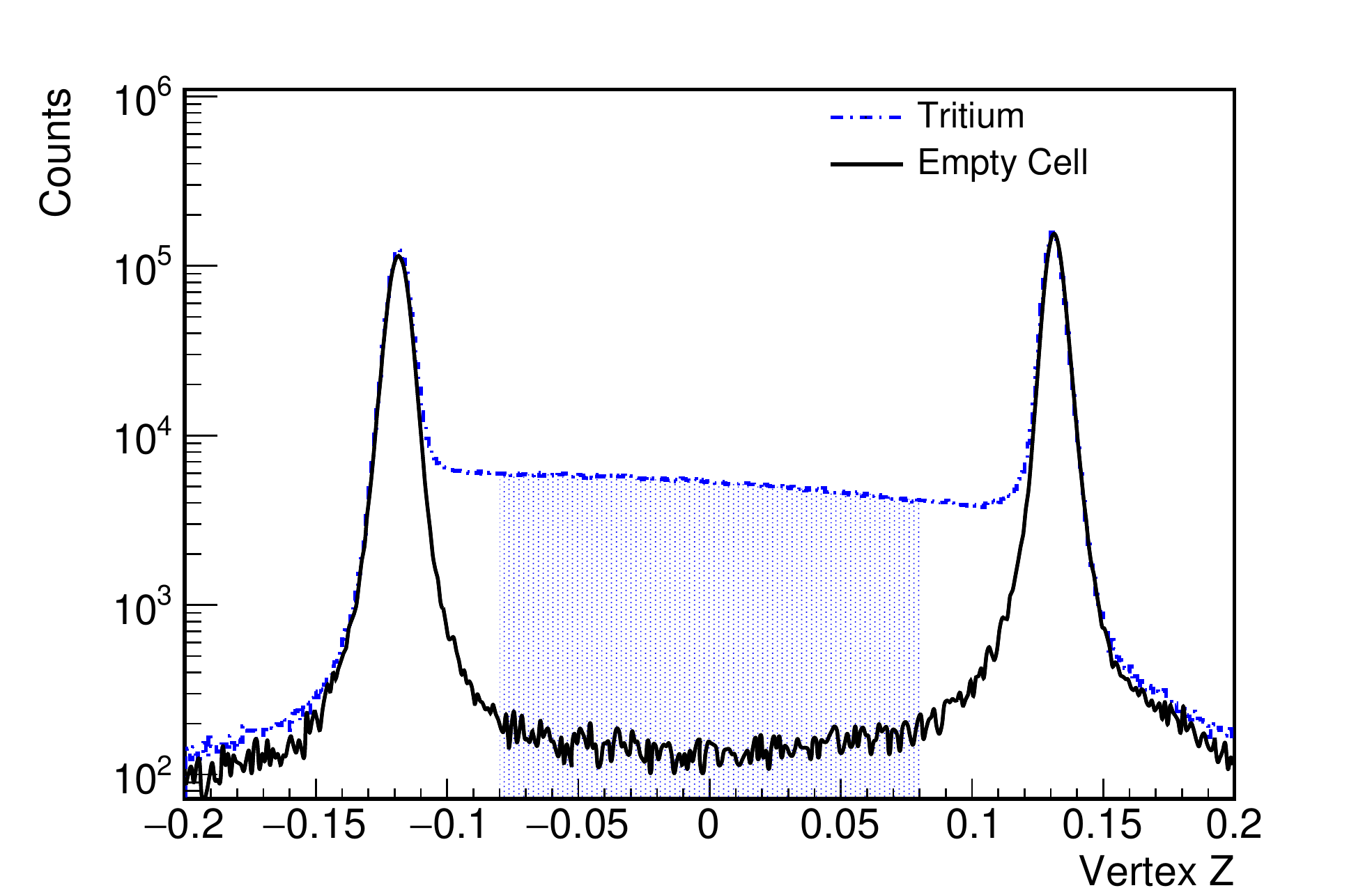}
    \flushleft
        \caption{\textbf{Target window contamination.} Number of events vs. position in the target along the beamline for the $^3$H cell (blue) and for the empty target (black) after scaling to the same luminosity as the target windows. The shaded region indicates the region used in the analysis.}\label{fig:endcap} 
\end{figure*}

\begin{figure*}[htb]
 \renewcommand\figurename{Extended Data Figure}
    \centering
        \includegraphics[width=89mm]{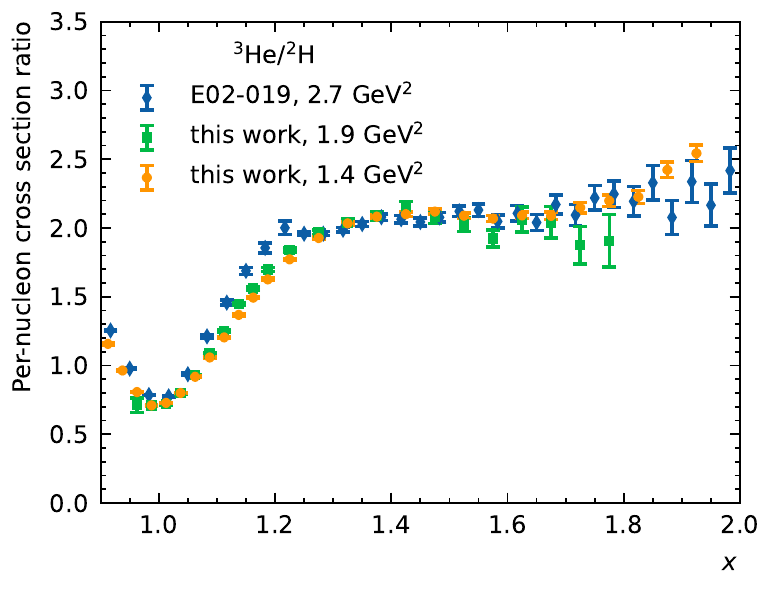}
    \flushleft
        \caption{\textbf{$^3$He/$^2$H per-nucleon cross section ratios} $^3$He/$^2$H ratio for this work and Ref.\cite{fomin12} are shown. Error bars show the combined statistical and uncorrelated systematic uncertainty (1$\sigma$ range); the normalization uncertainties are 1.18\% for this work, 1.8\% for E02-019.}\label{fig:nnratio_D}
\end{figure*}

\begin{table}[hb] 
\begin{center}
\small{\textbf{Extended Data Table 1 } $\vert$ 
\textbf{Cross section ratios at 17.00 degrees as shown in Fig.~\ref{fig:nnratio_3_x2}}}\\\vspace{5mm}
\renewcommand{\arraystretch}{3} 
\begin{tabular}{cccccc}
\hline
x     &  $\langle Q^2 \rangle$ (GeV$^2$) & $^3$H/$^3$He  & $^3$H/$^2$H  & $^3$He/$^2$H & isoscalar average\\\hline
0.6875 & 1.133 & 0.941$\pm$0.014 & 0.916$\pm$0.012 & 0.973$\pm$0.015 & 0.945 \\
0.7125 & 1.157 & 0.923$\pm$0.012 & 0.987$\pm$0.011 & 1.068$\pm$0.014 & 1.027 \\
0.7375 & 1.153 & 0.902$\pm$0.010 & 1.105$\pm$0.011 & 1.222$\pm$0.015 & 1.163 \\
0.7625 & 1.172 & 0.904$\pm$0.009 & 1.228$\pm$0.011 & 1.358$\pm$0.015 & 1.293 \\
0.7875 & 1.193 & 0.875$\pm$0.008 & 1.309$\pm$0.010 & 1.495$\pm$0.014 & 1.402 \\
0.8125 & 1.214 & 0.839$\pm$0.007 & 1.353$\pm$0.010 & 1.614$\pm$0.013 & 1.484 \\
0.8375 & 1.234 & 0.823$\pm$0.007 & 1.325$\pm$0.009 & 1.607$\pm$0.012 & 1.466 \\
0.8625 & 1.253 & 0.787$\pm$0.006 & 1.213$\pm$0.008 & 1.542$\pm$0.010 & 1.377 \\
0.8875 & 1.271 & 0.759$\pm$0.006 & 1.047$\pm$0.006 & 1.379$\pm$0.009 & 1.213 \\
0.9125 & 1.287 & 0.738$\pm$0.005 & 0.856$\pm$0.005 & 1.157$\pm$0.007 & 1.006 \\
0.9375 & 1.234 & 0.730$\pm$0.005 & 0.704$\pm$0.004 & 0.963$\pm$0.006 & 0.834 \\
0.9625 & 1.247 & 0.714$\pm$0.004 & 0.577$\pm$0.003 & 0.807$\pm$0.004 & 0.692 \\
0.9875 & 1.261 & 0.715$\pm$0.004 & 0.508$\pm$0.003 & 0.710$\pm$0.004 & 0.609 \\
1.0125 & 1.274 & 0.707$\pm$0.004 & 0.515$\pm$0.002 & 0.728$\pm$0.003 & 0.621 \\
1.0375 & 1.289 & 0.722$\pm$0.004 & 0.578$\pm$0.003 & 0.800$\pm$0.004 & 0.689 \\
1.0625 & 1.303 & 0.731$\pm$0.004 & 0.673$\pm$0.003 & 0.918$\pm$0.004 & 0.795 \\
1.0875 & 1.317 & 0.747$\pm$0.004 & 0.792$\pm$0.004 & 1.058$\pm$0.005 & 0.925 \\
1.1125 & 1.331 & 0.760$\pm$0.004 & 0.918$\pm$0.004 & 1.204$\pm$0.006 & 1.061 \\
1.1375 & 1.274 & 0.774$\pm$0.004 & 1.063$\pm$0.005 & 1.368$\pm$0.007 & 1.216 \\
1.1625 & 1.283 & 0.786$\pm$0.004 & 1.178$\pm$0.006 & 1.493$\pm$0.008 & 1.336 \\
1.1875 & 1.295 & 0.796$\pm$0.004 & 1.303$\pm$0.007 & 1.626$\pm$0.009 & 1.464 \\
1.2250 & 1.314 & 0.808$\pm$0.003 & 1.436$\pm$0.006 & 1.772$\pm$0.008 & 1.604 \\
1.2750 & 1.339 & 0.824$\pm$0.004 & 1.601$\pm$0.008 & 1.927$\pm$0.010 & 1.764 \\
1.3250 & 1.364 & 0.838$\pm$0.004 & 1.719$\pm$0.009 & 2.033$\pm$0.011 & 1.876 \\
1.3750 & 1.386 & 0.848$\pm$0.004 & 1.779$\pm$0.011 & 2.082$\pm$0.013 & 1.930 \\
1.4250 & 1.407 & 0.850$\pm$0.005 & 1.793$\pm$0.012 & 2.100$\pm$0.015 & 1.946 \\
1.4750 & 1.406 & 0.856$\pm$0.005 & 1.814$\pm$0.014 & 2.119$\pm$0.017 & 1.967 \\
1.5250 & 1.427 & 0.858$\pm$0.006 & 1.807$\pm$0.016 & 2.089$\pm$0.019 & 1.948 \\
1.5750 & 1.446 & 0.857$\pm$0.006 & 1.774$\pm$0.017 & 2.068$\pm$0.021 & 1.921 \\
1.6250 & 1.459 & 0.862$\pm$0.007 & 1.803$\pm$0.020 & 2.091$\pm$0.024 & 1.947 \\
1.6750 & 1.471 & 0.841$\pm$0.007 & 1.767$\pm$0.022 & 2.088$\pm$0.027 & 1.927 \\
1.7250 & 1.481 & 0.834$\pm$0.011 & 1.780$\pm$0.031 & 2.148$\pm$0.038 & 1.964 \\
1.7750 & 1.496 & 0.831$\pm$0.012 & 1.844$\pm$0.035 & 2.198$\pm$0.043 & 2.021 \\
1.8250 & 1.427 & 0.818$\pm$0.013 & 1.831$\pm$0.038 & 2.227$\pm$0.048 & 2.029 \\
1.8750 & 1.437 & 0.789$\pm$0.013 & 1.906$\pm$0.044 & 2.422$\pm$0.057 & 2.164 \\
1.9250 & 1.438 & 0.779$\pm$0.014 & 2.032$\pm$0.047 & 2.543$\pm$0.061 & 2.288 \\
1.9750 & 1.450 & 0.782$\pm$0.015 & 5.978$\pm$0.075 & 7.703$\pm$0.097 & 6.841 \\\hline
\end{tabular}
\caption{Kinematics and per-nucleon cross section ratios for the 17.00 degree ($Q^2 \approx 1.4$~GeV$^2$ in the SRC region) data with all uncorrelated uncertainties added in quadrature. The last column is the unweighted average of the $^3$He/$^2$H and $^3$H/$^2$H ratios. An additional normalization uncertainty of 0.78\% for $^3$H/$^2$H ratios and 1.18\% for $^3$He/$^3$H or $^3$He/$^2$H is not included.}\label{tab:data1}
\end{center}
\end{table}

\begin{table}[htb]
\begin{center}
\small{\textbf{Extended Data Table 2 } $\vert$ 
\textbf{ Cross section ratios at 20.88 degree as shown in Fig.~\ref{fig:nnratio_3_x2}}}\\\vspace{5mm}
\renewcommand{\arraystretch}{3.0}
\begin{tabular}{cccccc}
\hline
x     &  $\langle Q^2 \rangle$ (GeV$^2$) & $^3$H/$^3$He  & $^3$H/$^2$H  & $^3$He/$^2$H & isoscalar average\\\hline
0.9625 & 1.561 & 0.768$\pm$0.036 & 0.547$\pm$0.040 & 0.712$\pm$0.052 & 0.630 \\
0.9875 & 1.575 & 0.724$\pm$0.005 & 0.514$\pm$0.005 & 0.710$\pm$0.007 & 0.612 \\
1.0125 & 1.590 & 0.726$\pm$0.004 & 0.522$\pm$0.004 & 0.718$\pm$0.005 & 0.620 \\
1.0375 & 1.605 & 0.727$\pm$0.003 & 0.582$\pm$0.004 & 0.798$\pm$0.005 & 0.690 \\
1.0625 & 1.621 & 0.743$\pm$0.003 & 0.693$\pm$0.004 & 0.929$\pm$0.006 & 0.811 \\
1.0875 & 1.638 & 0.752$\pm$0.003 & 0.822$\pm$0.005 & 1.088$\pm$0.007 & 0.955 \\
1.1125 & 1.658 & 0.772$\pm$0.003 & 0.970$\pm$0.007 & 1.252$\pm$0.009 & 1.111 \\
1.1375 & 1.680 & 0.780$\pm$0.003 & 1.133$\pm$0.008 & 1.446$\pm$0.011 & 1.289 \\
1.1625 & 1.699 & 0.798$\pm$0.004 & 1.250$\pm$0.010 & 1.560$\pm$0.013 & 1.405 \\
1.1875 & 1.713 & 0.803$\pm$0.004 & 1.371$\pm$0.012 & 1.698$\pm$0.015 & 1.534 \\
1.2250 & 1.752 & 0.820$\pm$0.003 & 1.516$\pm$0.012 & 1.839$\pm$0.014 & 1.678 \\
1.2750 & 1.790 & 0.833$\pm$0.004 & 1.648$\pm$0.015 & 1.967$\pm$0.018 & 1.808 \\
1.3250 & 1.819 & 0.855$\pm$0.005 & 1.753$\pm$0.021 & 2.040$\pm$0.024 & 1.897 \\
1.3750 & 1.843 & 0.853$\pm$0.006 & 1.789$\pm$0.026 & 2.087$\pm$0.031 & 1.938 \\
1.4250 & 1.867 & 0.858$\pm$0.007 & 1.856$\pm$0.033 & 2.153$\pm$0.039 & 2.004 \\
1.4750 & 1.884 & 0.846$\pm$0.008 & 1.766$\pm$0.038 & 2.078$\pm$0.045 & 1.922 \\
1.5250 & 2.021 & 0.841$\pm$0.009 & 1.710$\pm$0.044 & 2.025$\pm$0.053 & 1.867 \\
1.5750 & 2.061 & 0.858$\pm$0.012 & 1.656$\pm$0.053 & 1.923$\pm$0.063 & 1.789 \\
1.6250 & 2.105 & 0.833$\pm$0.015 & 1.724$\pm$0.075 & 2.061$\pm$0.090 & 1.893 \\
1.6750 & 2.146 & 0.842$\pm$0.019 & 1.725$\pm$0.096 & 2.040$\pm$0.115 & 1.883 \\
1.7250 & 2.189 & 0.802$\pm$0.025 & 1.509$\pm$0.110 & 1.874$\pm$0.137 & 1.691 \\
1.7750 & 2.234 & 0.799$\pm$0.033 & 1.529$\pm$0.151 & 1.906$\pm$0.190 & 1.718 \\
1.8250 & 2.273 & 0.789$\pm$0.045 & 1.388$\pm$0.184 & 1.753$\pm$0.235 & 1.570 \\
1.8750 & 2.305 & 0.802$\pm$0.073 & 1.852$\pm$0.439 & 2.300$\pm$0.549 & 2.076 \\
1.9250 & 2.344 & 0.758$\pm$0.123 & 3.773$\pm$2.149 & 4.957$\pm$2.821 & 4.365 \\\hline
\end{tabular}
\caption{Kinematics and per-nucleon cross section ratios for the 20.88 degree ($Q^2 \approx 1.9$~GeV$^2$ in the SRC region) data with all uncorrelated uncertainties added in quadrature. The last column is the unweighted average of the $^3$He/$^2$H and $^3$H/$^2$H ratios. An additional normalization uncertainty of 0.78\% for $^3$H/$^2$H ratios and 1.18\% for $^3$He/$^3$H or $^3$He/$^2$H is not included.}\label{tab:data2}
\end{center}
\end{table}

\end{document}